\documentclass[prc,aps,floatfix,nofootinbib,twocolumn,superscriptaddress]{revtex4-2} 

\usepackage{graphicx}
\usepackage{color}
\usepackage{enumerate}
\usepackage{amssymb}
\usepackage{amsmath}
\usepackage{amsfonts}
\usepackage{bbm}
\usepackage{dsfont}
\usepackage{url}
\usepackage{braket}
\usepackage[normalem]{ulem}
\usepackage[colorlinks=true,allcolors=blue,breaklinks]{hyperref}

\renewcommand{\vec}[1]{\mbox{\boldmath $#1$}}

\begin{document}

\title{
Determination of nuclear deformations with an emulator for sub-barrier fusion reactions 
}

\author{Zehong Liao}
\affiliation{ 
Sino-French Institute of Nuclear Engineering and Technology, Sun Yat-sen University, Zhuhai 519082, China}
\affiliation{ 
Department of Physics, Kyoto University, Kyoto 606-8502,  Japan}

\author{K. Hagino}
\email[Corresponding author, ]{ hagino.kouichi.5m@kyoto-u.ac.jp}
\affiliation{ 
Department of Physics, Kyoto University, Kyoto 606-8502,  Japan} 
\affiliation{Institute for Liberal Arts and Sciences, Kyoto University, Kyoto 606-8501, Japan}
\affiliation{ 
RIKEN Nishina Center for Accelerator-based Science, RIKEN, Wako 351-0198, Japan
}

\author{Long Zhu}
\email[Corresponding author, ]{zhulong@mail.sysu.edu.cn}
\affiliation{ 
Sino-French Institute of Nuclear Engineering and Technology, Sun Yat-sen University, Zhuhai 519082, China} 

\author{S. Yoshida}
\affiliation{ 
School of Data Science and Management, Utsunomiya University, Mine, Utsunomiya 321-8505, Japan
}
\affiliation{ 
RIKEN Nishina Center for Accelerator-based Science, RIKEN, Wako 351-0198, Japan
}

\author{K. Uzawa}
\affiliation{ 
Nuclear Data Center, Japan Atomic Energy Agency, Tokai 319-1195,  Japan}


\begin{abstract}
Based on the eigenvector continuation, which is mathematically an instance of the
reduced basis method (RBM), 
we construct an emulator for coupled-channels calculations 
for heavy-ion fusion reactions at energies 
around the Coulomb barrier. 
We apply this to the $^{16}$O+$^{144,154}$Sm, $^{186}$W reactions and examine whether the emulator can 
be used to extract the deformation parameters of the target nuclei. 
We show that the emulator not only accelerates the calculations but also 
has an ability to accurately extract the nuclear shapes. 
This indicates that the emulator provides a powerful tool 
to systematically explore intrinsic shapes of atomic nuclei, 
enhancing our understanding of the fundamental properties of nuclear systems.
\end{abstract}

\maketitle


\section{Introduction}

A characteristic feature of atomic nuclei is a collective behavior originated from strong correlations 
among nucleons \cite{ring2004nuclear}. 
For instance, a collective rotational motion is emerged when a nucleus 
possesses an intrinsically deformed shape.
It is known that nuclei away from shell closures tend to be deformed in the ground state. 
The formation of rotational bands and a presence of large quadrupole electric moments serve as a 
direct experimental evidence for such nuclear deformations. 

Nuclear deformation plays a crucial role in low-energy nuclear reactions, e.g., heavy-ion fusion 
reactions at energies around the Coulomb barrier, for which the cross sections are significantly 
enhanced due to nuclear deformations \cite{10.1143/PTP.128.1061, PhysRevLett.67.3368, PhysRevC.47.R437, PhysRevC.52.3151, dasgupta1998measuring}. 
This enhancement of fusion cross sections arises from a distribution of barrier heights due to 
different orientations of a deformed nucleus. That is, 
passing through lower barriers in the distribution result in an increase of 
the overall penetration probability at energies below the original barrier.  
In this way, the static properties of nuclear deformation can be reflected in the observables 
of nuclear reactions. 

Notice that the picture of the barrier distribution is intimately related to 
the adiabatic approximation, that is, the orientation angle of the deformed target remains unchanged 
during the collision. 
This approximation is justifiable for fusion of medium heavy and heavy deformed nuclei. 
For example, for the $^{154}\text{Sm}$ nucleus, the energy of the first $2^+$ state ($0.082 \text{ MeV}$) is significantly lower than the characteristic energy scale of 
the $^{16}$O + $^{154}$Sm fusion reaction, e.g., the curvature of the Coulomb barrier, which is 
typically around 3.5 MeV. 
According to the uncertainty principle, this implies that the timescale of nuclear rotation is much 
longer than the characteristic time scale of 
the tunneling motion for the fusion process, and thus the rotational motion is frozen 
during the reaction. 

Incidentally, the nuclear shape dynamics has attracted a renewed interest in recent years, notably through ultra-relativistic heavy-ion collisions \cite{Collaboration_2025, PhysRevLett.124.202301, PhysRevLett.127.242301, PhysRevC.105.014905, PhysRevLett.128.022301,PhysRevLett.132.262301,star2024imaging, PhysRevLett.133.192301,PhysRevC.111.L011901,zymp-tyjj}. 
Unlike traditional methods that rely on external electromagnetic probes, 
this approach exploits the fact that these extremely energetic collisions 
generate a hot and dense quark–gluon plasma (QGP). The initial geometry of the colliding nuclei fundamentally determines the initial spatial distribution of the QGP, which governs the subsequent evolution. 
By analyzing the momentum-space observables of the final-state particles, 
one could effectively trace back the reaction dynamics to reconstruct the 
initial nuclear configuration which reflect the nuclear shapes. 
In such analysis, the adiabatic approximation is fully justified, and thus 
the ultra-relativistic heavy-ion collisions exhibit a large similarity to low-energy heavy-ion 
fusion reactions regarding the treatment of nuclear shape dynamics \cite{HaginoKitazawa2025}. 

When the rotational energies are not small, e.g., for light nuclei, the adiabatic 
approximation breaks down, and one has to solve the coupled-channels equation as they are \cite{10.1143/PTP.128.1061,HAGINO2022103951}. 
Using such approach, the quadrupole and hexadecapole deformation parameters of $^{24}$Mg and $^{28}$Si have been 
recently extracted from the 
quasi-elastic scattering of $^{24}\mathrm{Mg} + {}^{90}\mathrm{Zr}$ and 
$^{28}\mathrm{Si} + {}^{90}\mathrm{Zr}$ systems with high precision \cite{GUPTA2020135473,GUPTA2023138120}. 
The optimum values of the deformation parameters are consistent with theoretical predictions of nuclear structure calculations 
as well as with the previous experimental determinations. 
However, a major computational bottleneck arises in such analyses. That is, 
a large number of coupled-channels calculations have to be repeated as 
the nuclear deformation parameters 
must be iteratively varied, that substantially increase the computational cost. 

The computational burden can be alleviated by using emulators, or surrogate models, which accurately approximate the original calculations with much cheaper computational costs \cite{RevModPhys.94.031003,Dobaczewski_2014}. 
Broadly speaking, emulators can be classified into data-driven approaches, model-driven approaches, or a hybrid of these \cite{692f0404,Brunton_Kutz_2019}. Data-driven emulators, such as Gaussian processes, typically construct a mapping relationship between inputs and outputs. They are 
generally non-intrusive but are particularly effective in analyzing extensive experimental data 
sets and supporting theoretical predictions \cite{PhysRevC.101.044307, Alhassan:2020zlx, fang_bayesian_2024}. 
In contrast, model-driven emulators are physics-based and utilize the exact governing equations of the system, from which they derive the reduced-order equations\cite{melendez_model_2022}. Within the broader computational science and engineering communities, such model-driven techniques have been extensively developed under the framework of the Reduced Basis Method (RBM) \cite{ChenChen2017, quarteroni2016reduced, PhysRevC.106.054322, Drischler:2022ipa, melendez_model_2022}. In the nuclear physics community, the Eigenvector Continuation (EC)
exemplifies a specific application of the RBM  mathematical framework \cite{PhysRevLett.121.032501,RevModPhys.96.031002}. In this application, approximate solutions to a model Hamiltonian are constructed by linearly superposing a small 
set of its eigenvectors computed for selected parameter values. This method has been widely applied to uncertainty quantifications and parameter sensitivity analyses \cite{PhysRevLett.123.252501, KONIG2020135814, PhysRevC.104.064001, PhysRevC.110.014309, 4cnl-5dnm, 10.1093/ptep/ptac057, PhysRevC.106.054322}, 
especially for cases where the underlying model computations are extremely time-consuming. The 
method has been demonstrated to be highly effective for nuclear structure problems \cite{PhysRevC.107.064316, franzke2025hartreefockemulatorsnucleiapplication, PhysRevC.111.064318, 4ccs-66c6} as well as  nuclear reaction problems \cite{FURNSTAHL2020135719, DRISCHLER2021136777, MELENDEZ2021136608, PhysRevC.103.014612, PhysRevC.106.024611, PhysRevC.105.064004, LIU2024139070, PhysRevC.109.044612,catacorarios2025,Jin2026}. 

In Ref. \cite{zpsz-7jld}, we constructed and tested an emulator for 
a simple one-dimensional coupled-channels model. To this end, we employed the discrete basis formalism together 
with the Kohn variational principle \cite{PhysRevC.110.054610}. 
In this paper, we extend the emulator to more realistic scattering problems 
in three dimensions \cite{Hagino:1999xb, 10.1143/PTP.128.1061, HAGINO2022103951}. 
We particularly analyze the 
$^{16}$O+$^{144,154}$Sm, $^{186}$W fusion reactions at subbarrier energies, and examine whether the 
emulator can be utilized to extract deformation parameters from the experimental fusion cross sections. 

The paper is organized as follows. 
In Sec. \ref{sec2}, we will introduce the emulator for coupled-channels calculations based on the {\tt CCFULL} 
model \cite{Hagino:1999xb}. We will  
describe in details the workflow of extracting the nuclear shape from 
experimental fusion cross sections with the emulator. We will then apply the emulator 
to the $^{16}$O+$^{144,154}$Sm, $^{186}$W reactions in 
Sec. \ref{sec3} to extract the deformation parameters of the 
target nuclei. We will also discuss the performance of the emulator, that is, the computation speed and the 
accuracy. 
We will then summarize the paper in Sec. \ref{sec4}.

\section{METHODS}\label{sec2}

\subsection{Workflow}\label{workflow}

\begin{figure*}[htbp]
    \centering
        \includegraphics[width=17.0cm]{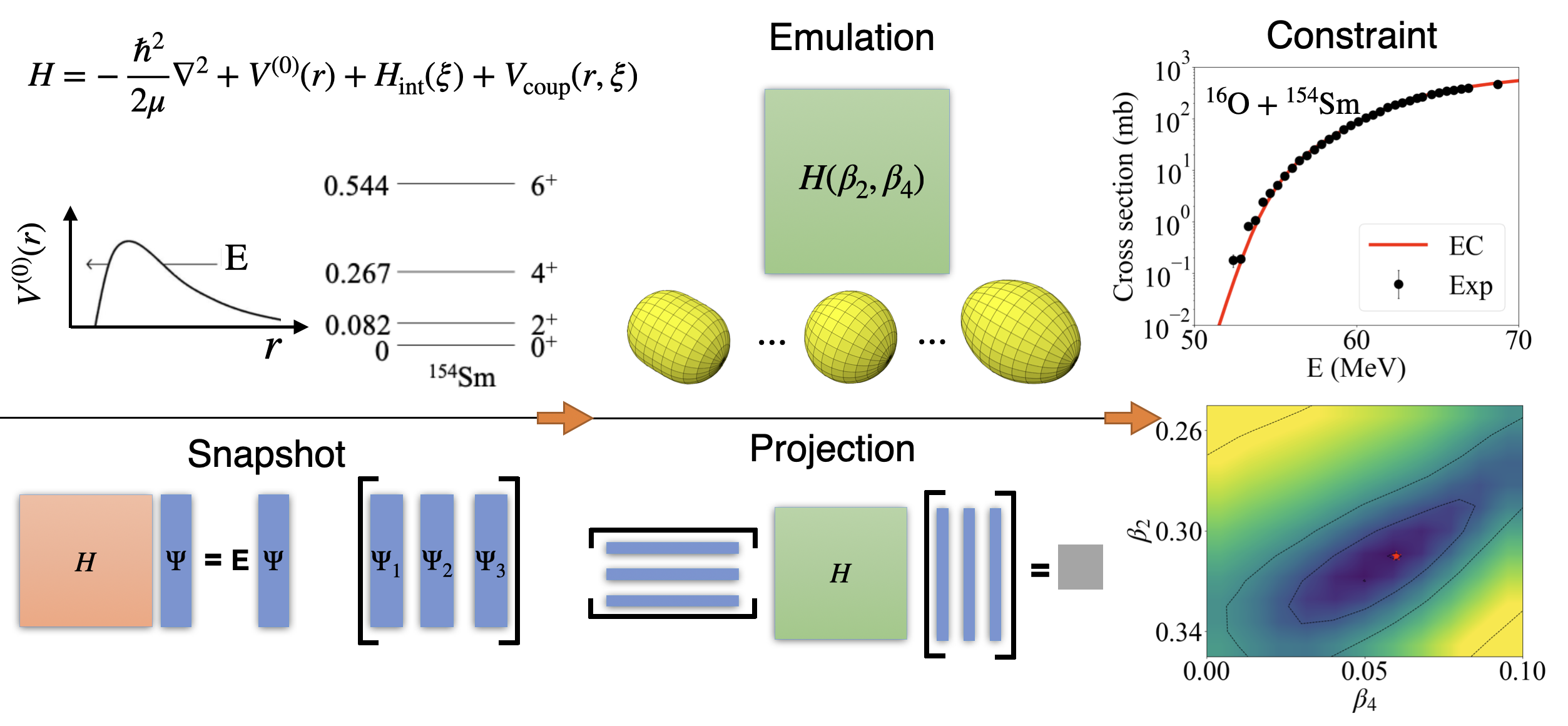}
    \caption{The workflow of the emulator for determining the nuclear shape.
    High-fidelity eigen-functions $\Psi$ of the coupled-channels equations are first obtained 
    for a selected set of parameters (Snapshot). 
    The full Hamiltonian $H$ is projected onto the subspace formed by these eigen-functions (Projection) 
    to build an emulator, which facilitates rapid predictions of fusion cross sections (Emulation). 
    The results of the emulator are then compared with experimental data to 
    perform a $\chi^2$ analysis, with which optimum values of the parameters are extracted (Constraint). 
    } \label{Pic_1}
\end{figure*}

In this paper, we employ the EC method as a surrogate model to efficiently compute fusion 
cross sections under different sets of model parameters. 
To this end, we use the coupled-channels code {\tt CCFULL}\cite{Hagino:1999xb}
to obtain physical observables. 
As we will show, the EC emulator enables accurate reproductions of the coupled-channels calculations 
while significantly reducing the computational cost, thereby providing a flexible tool for systematic parameter exploration. The basic procedure is as follows. 
Firstly, a high-fidelity snapshot is taken by solving the full high-dimensional Hamiltonian for a few selected parameter sets (that is, training points), yielding the scattering wave functions $\{\Psi_i\}$. 
Secondly, these solutions are used to construct a low-dimensional EC subspace. 
Via the projection, one can construct an effective emulator across the entire parameter space. 
The resultant approximate wave function is given as a liner superposition of $\{\Psi_i\}$. 
Finally, the emulator is used to compute observables, that is, fusion cross sections. The calculated results are compared to experimental data to map the $\chi^2$ landscape, which is then used to identify the optimal parameters 
that define the intrinsic nuclear shape. This process is illustrated in Fig. \ref{Pic_1}. 
By providing fast calculations, this approach not only enhances the predictive power of the model 
but also provides novel insights into the role of nuclear structure in heavy-ion fusion reactions.

We will explain the details of each piece of the calculations in the following subsections. 

\subsection{Coupled-channels formalism}\label{subsec:CCF}

In heavy-ion fusion reactions, the collision process is inherently dynamic, involving strong interplay between the relative motion and the intrinsic degrees of freedom of the colliding nuclei. 
The total fusion probability is significantly affected by such couplings, 
especially at energies near and below the Coulomb barrier. 
Consequently, it is essential for a quantitative description of fusion 
to accurately incorporate 
the channel coupling effects. 
This necessitates employing a reaction Hamiltonian that explicitly accounts for these internal 
structures. Such Hamiltonian can be generally written as \cite{Hagino:1999xb, 10.1143/PTP.128.1061, HAGINO2022103951}:
\begin{equation}
    H = -\frac{\hbar^{2}}{2\mu}\vec{\nabla}^2+ V^{(0)}(r)+H_{\mathrm{int}}(\xi) + V_{\mathrm{coup}}(r, \xi),
    \label{eq:totH}
\end{equation}
where $r$ denotes the coordinate of the relative motion between the target and the projectile nuclei, and $\mu$ is the reduced mass of the reaction system. $V^{(0)}(r)$ is the bare potential, which is the sum of the Coulomb and the nuclear potentials. $H_{\mathrm{int}}(\xi)$ describes the intrinsic Hamiltonian that governs the excitation spectrum of the colliding nuclei, where $\xi$ represents the internal degrees of freedom. Finally, $V_{\mathrm{coup}}(r, \xi)$ represents the coupling potential between these intrinsic 
excitations and the relative motion. In conventional coupled-channels calculations, 
the coupling potential $V_{\mathrm{coup}}$ is typically constructed based on the geometrical collective model. 

For each total angular momentum $J$, the total wave function $\Psi(r,\xi)$ is expanded as 
\begin{equation}
    \Psi(r,\xi)=\sum_n\frac{\phi_n(r)}{r}|n(\xi)\rangle,
    \label{eq:cc-wf}
\end{equation}
where $|n(\xi)\rangle$ is the eigen-functions of the intrinsic 
Hamiltonian $H_{\mathrm{int}}(\xi)$, satisfying $H_{\mathrm{int}}(\xi)|n\rangle=\epsilon_n|n\rangle$. 
Here $\epsilon_n$ is the eigen-energy of the state $|n(\xi)\rangle$. 
To write down Eq. (\ref{eq:cc-wf}), we have used the iso-centrifugal approximation and ignored the spin of the 
intrinsic state $|n(\xi)\rangle$ \cite{Hagino:1999xb, 10.1143/PTP.128.1061, HAGINO2022103951}. 
The coupled-channels equations for $\phi_n(r)$ are obtained as 
\begin{eqnarray}
&&\left[-\frac{\hbar^2}{2\mu}\frac{d^2}{dr^2}+\frac{J(J+1)\hbar^2}{2\mu r^2}
+V^{(0)}(r)-E+\epsilon_n\right]\phi_n(r) \nonumber \\
&&=-\sum_m\langle n|V_{\rm coup}(r)|m\rangle\phi_m(r)\equiv -\sum_m V_{nm}(r)\phi_m(r), 
\label{eq:cc}
\end{eqnarray}
where $E$ is the incident energy in the center of mass frame. 
The coupled-channels equations are solved by imposing the incoming wave boundary condition (IWBC), 
that is, there are only incoming waves at $r \leq r_\mathrm{min}$ 
\cite{Hagino:1999xb, 10.1143/PTP.128.1061, HAGINO2022103951}, for which 
$r_\mathrm{min}$ is taken to be the minimum position of the Coulomb pocket inside the barrier, as in 
the {\tt CCFULL} code \cite{Hagino:1999xb}. 
In addition, one imposes the boundary condition that 
there are only the outgoing waves at infinity for all the channels except for the entrance channel ($n$ = 1), 
which has the incoming wave as well. 
These boundary conditions read, 
\begin{eqnarray}
    \phi_n(r)&\to& H^{(-)}_{J}(k_{n}r)\,\delta_{n,1}+R^{(J)}_n H^{(+)}_{J}(k_{n}r), ~(r \to \infty), \nonumber \\
    \label{eq:bc1-2}
    \\
    &\to&T^{(J)}_n \exp(-ik_{n}(r) r),  ~~~~~~~~~~~~~~~~~~(r= r_{\mathrm{min}}), \nonumber \\
    \label{eq:bc2-2}
\end{eqnarray}
where $k_n(r)$ is the local wave number for the $n$-th channel with $k_n=k_n(r=\infty)$, 
and $T_n$ and $R_n$ are the transmission and the reflection coefficients for the channel $n$, respectively. 
$H^{(-)}_{J}$ and $H^{(+)}_{J}$ in Eq. (\ref{eq:bc1-2}) are the incoming and the outgoing Coulomb wave 
functions, respectively. 
The fusion cross sections $\sigma_{\mathrm{fus}}(E)$ are then calculated as 
\begin{equation}\label{eq:crosssection}
    \sigma_{\mathrm{fus}}(E) = \frac{\pi}{k^2}\sum_{J=0}^\infty(2J+1)P_{J}(E), 
\end{equation}
where $k\equiv k_1$ and $P_J(E)$ is the penetrability given by 
\begin{equation}
    P_{J}(E)=\sum_n \frac{k_n(r_{\rm min})}{k}\,|T_n^{(J)}|^2=
    1-\sum_n \frac{k_n}{k}\,|R_n^{(J)}|^2.
\end{equation}

We use a Woods-Saxon parameterization for 
the nuclear potential, 
\begin{equation}
    V_N^{(0)}(r) = - \frac{V_{0}}{1 + \exp[(r - R_{0})/a]}, 
    \label{eq:WS}
\end{equation}
where $V_0$ and $a$ are the depth and the diffuseness parameters, respectively. 
$R_{0} = r_{0}(A^{1/3}_{P} + A^{1/3}_{T})$ is the sum of the target and the projectile radii, 
in which $A_P$ and $A_T$ are the mass numbers of the projectile and the target nuclei, respectively.
For a deformed target nucleus, the coupling arises from the rotational excitations of the nucleus 
associated with the nuclear deformation. 
For the axially symmetric quadrupole ($\lambda=2$) and hexadecapole ($\lambda=4$) deformations, 
the radius of the target nucleus, $R_T$, is modified as, 
\begin{equation}
    R_T \rightarrow R_T + \hat{O} = R_T + \beta_2 R_T Y_{20}(\theta) + \beta_4 R_T Y_{40}(\theta),
\end{equation}
where $R_T$ is parameterized as $R_T=r_{\mathrm{coup} }A_{T}^{1/3}$, and $\beta_2$ and $\beta_4$ are the quadrupole ($\lambda=2$) and the hexadecapole ($\lambda=4$) deformation parameters, respectively. 
$Y_{\lambda 0}(\theta)$ are the spherical harmonics. Notice that the angle $\theta$ acts as 
the intrinsic coordinate $\xi$ in Eq. (\ref{eq:totH}). 
The Woods-Saxon potential, Eq. (\ref{eq:WS}), is then modified as, 
\begin{equation}\label{eq:Potenital_nuclear}
    V^{(N)}(r, \hat{O}) = - \frac{V_{0}}{1 + \exp[(r - R_{0} - \hat{O})/a]}, 
\end{equation}
For the rotational excitations of the ground rotational band of the target nucleus, 
the coupling matrix elements $V_{nm}^{(N)}(r)=\langle n|V^{(N)}(r, \hat{O})|m\rangle$ 
are calculated between the intrinsic states $|n\rangle = |I0\rangle$ and $|m\rangle = |I'0\rangle$, 
where $I$ and $I'$ are the spin of the rotational states. 
These matrix elements can be efficiently and accurately obtained using a matrix algebra \cite{PhysRevC.48.2326}. Within this algebraic framework, the first step is to diagonalize the deformation operator $\hat{O}=\beta_2 R_T Y_{20}(\theta) + \beta_4 R_T Y_{40}(\theta)$ to obtain the eigenvalues $\lambda_\alpha$ and the corresponding eigenvectors $|\alpha\rangle$ that satisfy,
\begin{equation}
    \hat{O}\ket{\alpha} = \lambda_{\alpha}\ket{\alpha}.
\end{equation}
For the rotational coupling, the matrix elements of $\hat{O}$ read \cite{Hagino:1999xb}, 
\begin{align}
\hat{O}_{nm} 
&= \beta_2 R_T \sqrt{\frac{5(2I + 1)(2I^{\prime} + 1)}{4\pi}} 
\left(
\begin{array}{ccc}
I & 2 & I^{\prime} \\
0 & 0 & 0
\end{array}
\right)^2 \nonumber \\ &
+ \beta_4 R_T \sqrt{\frac{9(2I + 1)(2I^{\prime} + 1)}{4\pi}} 
\left(
\begin{array}{ccc}
I & 4 & I^{\prime} \\
0 & 0 & 0
\end{array}
\right)^2.
\end{align}
The nuclear coupling matrix elements are then evaluated as, 
\begin{equation}
 \begin{aligned}
V_{nm}^{\left(N\right)}(r) = \sum_{\alpha}\left\langle I0|\alpha\right\rangle\left\langle\alpha|I^{\prime}0\right\rangle V_{N}\left(r,\lambda_{\alpha}\right)- V^{(0)}_N(r)\delta_{n,m},
\end{aligned}   
\end{equation}
The last term is included to prevent a double-counting of the diagonal component of the bare nuclear potential. 

For the Coulomb coupling, $V^{(C)}_{\mathrm{coup}}(r, \theta)$, 
we consider the multipole expansion of the Coulomb potential 
up to the second order in $\beta_2$ and up to the first order in $\beta_4$ \cite{Hagino:1999xb}. 
This leads to, 
\begin{equation}
\begin{aligned}
V_{nm}^{(C)}(r) 
&= \frac{3 Z_{P} Z_{T} e^2}{5} \frac{R_{T}^{2}}{r^{3}} \sqrt{ \frac{5 (2I +1)(2I' +1)}{4\pi} } 
\\ & \times\left( \beta_{2} + \frac{2}{7} \sqrt{ \frac{5}{\pi} } \beta_{2}^{2} \right)
\begin{pmatrix}
I & 2 & I' \\
0 & 0 & 0
\end{pmatrix}^{2} 
\\
&\quad + \frac{3 Z_{P} Z_{T} e^2}{9} \frac{R_{T}^{4}}{r^{5}} \sqrt{ \frac{9 (2I +1)(2I' +1)}{4\pi} } 
\\ & \times \left( \beta_{4} + \frac{9}{7} \sqrt{ \frac{1}{\pi} } \beta_{2}^{2} \right)
\begin{pmatrix}
I & 4 & I' \\
0 & 0 & 0
\end{pmatrix}^{2},
\end{aligned}
\end{equation}
where $Z_P$ and $Z_T$ are the atomic numbers of the projectile and the target nuclei, respectively. 

The coupled-channels formalism can be applied not only to the rotational couplings of deformed 
nuclei but also to the vibrational couplings of spherical nuclei. 
The actual formulas for the matrix elements for the vibrational couplings can be found in Ref. \cite{Hagino:1999xb}.

\subsection{Discrete basis method}\label{subsec:DB}

The discrete basis method (DBM) generally refers to solving quantum mechanical problems 
by expanding a wave function $\Psi(r)$ with a set of discrete, square-integrable basis functions. 
A general form of the wave function $\Psi(r)$ reads 
\begin{eqnarray}
    \Psi = \sum^{N}_{i=1} b_{i} \phi_{i} + c\phi_{H}, 
\end{eqnarray}
where $\phi_{i}$ are basis functions, chosen to form a complete (or approximately complete) set within the region of interest, and $b_{i}$ are the corresponding coefficients to be determined. 
The last term may be added to satisfy the boundary condition for $\Psi$. 
In the generator coordinate method (GCM), $\Psi$ is a many-body wave function of a nucleus, while 
$\phi_{i}$ may be Slater determinants specified with quadrupole deformations $\{\beta_i\}$ \cite{ring2004nuclear}. 
For a scattering problem, on the other hand, $\Psi$ is a scattering wave function while 
$\phi_{i}$ may represent a discretized coordinate, $\phi_{i}=\delta_{i,r_i}$ \cite{PhysRevC.109.034611}. 
In this formalism, the scattering problem is transformed into a matrix problem. 

The discrete basis formalism has been successfully applied to induced fission reactions \cite{BERTSCH201968,ALHASSID2020168233,PhysRevC.98.014604} as well as to barrier-penetration problems \cite{PhysRevC.109.054606,PhysRevC.109.034611}. Motivated by this success, a new scattering 
theory based on the DBM combined with the Kohn variational principle was recently proposed in Ref. \cite{PhysRevC.110.054610}. 
The primary aim of this paper is to apply this new formalism to construct an emulator for multi-channel 
scattering using the EC technique. To this end, we extend our previous analysis of single-channel and two-channel problems with a one-dimensional Gaussian potential \cite{zpsz-7jld} 
to a more practical and physical problem of heavy-ion fusion reactions in three-dimensinaol space as modeled by the computer code {\tt CCFULL} \cite{Hagino:1999xb}. 

The discrete basis formalism can be implemented into the coupled-channels method as follows. 
For the $N_{\mathrm{ch}}$ channel problem, we first discretize the coordinate $r$ 
into $N$ mesh points with the mesh spacing $\Delta r$, that is, 
from $r_1\equiv r_\mathrm{min}$ to $r_{N} \equiv r_\mathrm{max}
= r_1+(N-1)\Delta r$.  
By applying the three-point finite difference formula for the second derivative, 
the coupled-channels equations (\ref{eq:cc}) are transformed into the matrix representation as
\begin{equation}
    \sum_{n'}\sum_{i'} (H_{ni,n'i'}-E\delta_{n,n'}\delta_{i,i'})\phi_{n'i'}=0, 
    \label{eq:cc-disc}
\end{equation}
where $n$ and $i$ refer to the channels and the coordinate $r_{i} = r_{\mathrm{min}} + (i-1)\Delta r$, 
respectively, and $\phi_{n'i'}$ represents the wave function $\phi_n(r_i)$. 
The matrix $H_{ni,n'i'}$ in Eq. (\ref{eq:cc-disc}) is given by, 
\begin{eqnarray}\label{eq:H}
H_{ni,n'i'}&=&\delta_{n,n'}(-t\,\delta_{i,i'+1} +2t\,\delta_{i,i'}-t\,\delta_{i,i'-1}) \nonumber \\
&&+\delta_{n,n'}\delta_{i,i'}\left(\frac{J(J+1)\hbar^2}{2\mu r_i^2}
+V^{(0)}(r_i)+\epsilon_n\right) \nonumber \\
&&+\delta_{i,i'}V_{nn'}(r_i), 
\end{eqnarray}
where $t$ is defined as $t\equiv \frac{\hbar^2}{2\mu\Delta r^2}$. 
Notice that the boundary condition Eq. (\ref{eq:bc2-2}) can be easily implemented by changing 
$H_{n1,n1}$ to 
$H_{n1,n1}\to H_{n1,n1}-te^{ik_n(r_{\rm min})\Delta r}$, which follows from 
the boundary condition of $ \phi_{n,0} = \phi_{n,1} \exp(ik_n(r_{\rm min})\Delta r)$. 
After this modification, we define the 
$N_{\mathrm{ch}}(N+1) \times N_{\mathrm{ch}}(N + 2)$ dimension Hamiltonian matrix $\tilde{H}$ 
defined as 
\begin{equation}
\tilde{H}_{ni,n'i'}=H_{ni,n'i'}-E\delta_{n,n'}\delta_{i,i'}
-te^{ik_n(r_{\rm min})\Delta r}\delta_{n,n'}\delta_{i,i'}\delta_{i,1}, 
\label{eq:H-disc}
\end{equation}
where $i$ and $i'$ run from 1 to $N+1$ and from 1 to $N+2$, respectively, and both $n$ and $n'$ run 
from 1 to $N_{\rm ch}$. 

To implement the Kohn variational principle, 
we define the basis functions as 
\begin{eqnarray}
\psi_{sj}(n'i')&=&\delta_{i',j}\delta_{n',s},~~~(1\leq j \leq N), \\
\psi_{s,N+1}(n'i')&=&\delta_{i',N+1}\delta_{n',s} H^{(+)}_{J}(k_n(r_{\mathrm{max}}+\Delta r)) \nonumber \\
&&+\delta_{i',N+2}\delta_{n',s} H^{(+)}_{J}(k_n(r_{\mathrm{max}}+2\Delta r), 
\label{eq:psi_extra}
\\
\psi^{(-)}(n'i') &=& \delta_{n',1}(\psi_{n',N+1}(n'i'))^*,
\label{eq:psi_extra2}
\end{eqnarray}
where $i'$ runs from 1 to $N+2$ while $n'$ runs  
from 1 to $N_{\rm ch}$ as in Eq. (\ref{eq:H-disc}). 
We then 
expand the total wave function with these basis functions as 
\begin{equation}
    \Psi(n'i') = \sum_{s = 1} ^{N_{\mathrm{ch}}}\sum^{N+1}_{j=1} b_{sj} \psi_{sj}(n'i') + \psi^{(-)}(n'i').
    \label{eq:discrete-basis}
\end{equation}
Notice that this form guarantees the boundary condition (\ref{eq:bc1-2}). 
By imposing $\tilde{H}'\Psi=0$, one obtains 
\begin{eqnarray}
&&\sum_{s = 1} ^{N_{\mathrm{ch}}}\sum^{N+1}_{j=1} 
\left(\sum_{n' = 1} ^{N_{\mathrm{ch}}}\sum^{N+2}_{i'=1}
\tilde{H}_{ni,n'i'}
\psi_{sj}(n'i')\right) b_{sj} \nonumber \\
&&+ 
\sum_{n' = 1} ^{N_{\mathrm{ch}}}\sum^{N+2}_{i'=1}
\tilde{H}_{ni,n'i'}
\psi^{(-)}(n'i')=0,
\end{eqnarray}
or
\begin{equation}
\sum_{s = 1} ^{N_{\mathrm{ch}}}\sum^{N+1}_{j=1} 
(\tilde{H}\psi)_{ni,sj}b_{sj} =-
(\tilde{H}\psi^{(-)})_{ni}\equiv -\tilde{\psi}_{ni},
\end{equation}
from which the amplitudes $\{b_{ni}\}$ can be obtained as
\begin{equation}
b_{ni}=-\sum_{s=1}^{N_{\mathrm{ch}}}\sum_{j=1}^{N+1}\left[(\tilde{H}\psi)^{-1}\right]_{ni,sj}\tilde{\psi}_{sj}.
\end{equation}
The penetrability is then computed as,
\begin{equation}\label{eq:Pexact}
    P_{J}(E)
    =\sum_{n=1}^{N_{\mathrm{ch}}}\frac{k_n(r_{\rm min})}{k}|b_{n1}|^2=
    1-\sum_{n=1}^{N_{\mathrm{ch}}}\frac{k_n}{k}|b_{n,N}|^2. 
\end{equation}

It is worth noting that replacing the standard three-point 
difference formula in the Hamiltonian matrix with the Numerov 
formula significantly enhances the stability and the convergence of the calculated wave functions. 
In the actual calculations, we shall employ this version of the discrete basis formalism. See Appendix. \ref{appendix:A} for details. Besides, employing the Lagrange mesh basis \cite{BAYE20151, DESCOUVEMONT2016199, PhysRevC.106.024611} rather than the equally spaced mesh representation in the discrete-basis formalism may be an avenue for future integrations, as it would effectively reduce the size of the model space and further reduce the computation time.

\subsection{Eigenvector continuation}\label{subsec:EC}

The discrete basis formalism presented in the previous subsection 
provides a natural starting point to implement the EC method into a construction 
of an emulator for coupled-channels calculations. 
The coupling potential, $V_{\rm coup}$, inherently depends on a set of physical parameters, 
such as the nuclear deformation parameters $\beta_{2}$ and $\beta_{4}$, 
as mentioned in Sec. \ref{subsec:CCF}. 
For a given set of those parameters, $\{\beta^{(i)}_\lambda\}$, we solve the coupled-channels equations and 
construct a vector as, 
\begin{equation}
\begin{aligned}
            \phi_i 
&= \sum_{s = 1} ^{N_{\mathrm{ch}}}\sum^{N}_{j=1} b^{(i)}_{sj} \psi_{sj} \\
                &= \{b^{(i)}_{11}, b^{(i)}_{21}, \cdots,b^{i)}_{N_{\rm ch}1},b^{(i)}_{21}, b^{(i)}_{22},\cdots, b^{(i)}_{N_{\rm ch}N} , 0, \cdots, 0\}^{T},
\end{aligned}
\end{equation}
where $T$ represents the transpose. That is, $\phi_i$ is the same as $\Psi$ in Eq. (\ref{eq:discrete-basis}), except that 
the $\psi_{s,N+1}$ and the $\psi^{(-)}$ components defined by Eqs. (\ref{eq:psi_extra}) and (\ref{eq:psi_extra2}), respectively, 
are removed from there. 
Note that the dimension of the vector $\phi_i$ is still $N_{\rm ch}(N+2)$. 
Following the philosophy of EC, we construct similar functions $\phi_{i}$ for 
a few selected parameter sets (that is, the training points), and superpose 
them to construct the total wave function as

\begin{equation}
\begin{aligned}
    \Psi_{\mathrm{EC}} &= \sum_{i=1}^{N_{\mathrm{EC}}}c_{i}\phi_{i} +  \sum_{s=1}^{N_{\mathrm{ch}}}c_{0s}\psi_{s,N+1} +\psi^{(-)}, 
\\ &=  \sum_{i=1}^{N_{\mathrm{EC}} + N_{\mathrm{ch}}}c_{i}\phi_{i} + \psi^{(-)}, 
\end{aligned}
\end{equation}
where $N_{\mathrm{EC}}$ is the number of basis vectors for EC, and 
$c_{s}$ and  $\phi_{s}$ for $s>N_{\rm EC}$ are defined as $c_s\equiv c_{0s}$ and 
$\phi_{s} \equiv \psi_{s,N+1}$, respectively. 
From the condition of $\tilde{H}\Psi_{\mathrm{EC}}=0$, one obtains 
\begin{equation}
 \sum_{j=1}^{N_{\mathrm{EC}} + N_{\mathrm{ch}}}\langle\phi'_i|\tilde{H}|\phi_j\rangle c_j=-\langle\phi'_i|\tilde{H}|\psi^{(-)}\rangle\equiv -d_i,
\end{equation}
where $\phi'_i$ and $d_i$ are defined as $|\phi'_i\rangle\equiv\tilde{H}|\phi_i\rangle$ and $d_i\equiv\langle\phi'_i|\tilde{H}|\psi^{(-)}\rangle$, respectively. The coefficients $c_i$ can then be obtained as 
\begin{equation}
c_i=-\sum_{i=1}^{N_{\mathrm{EC}} + N_{\mathrm{ch}}}(A^{-1})_{ij}d_j, 
\end{equation}
with $A_{ij}\equiv\langle\phi'_i|\tilde{H}'|\phi_j\rangle$, from which 
the penetrability is computed as
\begin{equation}\label{ec:P_ec}
    P_J = \sum^{N_{\mathrm{ch}}}_{s=1} \frac{k_{s}(r_{\rm min})}{k} |c_sb_{s1}|^2 
= 1-\sum^{N_{\mathrm{ch}}}_{s=1} \frac{k_{s}}{k} |c_{s+N_{\rm EC}}|^2.
\end{equation}

\section{RESULTS AND DISCUSSIONS}\label{sec3}

\subsection{Octupole vibration of $^{144}\mathrm{Sm}$}

Let us now apply the EC emulator to realistic nuclear fusion reactions. 
We first consider the $^{16}$O+$^{144}$Sm reaction. 
The target nucleus, $^{144}\text{Sm}$, has the neutron number $N$=82, and is regarded as a 
spherical nucleus. The first 3$^-$ state at 1.81 MeV is interpreted as 
a collective octupole vibrational state. From the measured $B(E\lambda)$ value, the 
(dynamical) deformation parameter $\beta_3$ is estimated to 
be $\beta_3 = 0.21$ \cite{PhysRevC.52.3151,PhysRevLett.79.2943}. 

\begin{figure}[htbp]
    \centering
        \includegraphics[width=8cm]{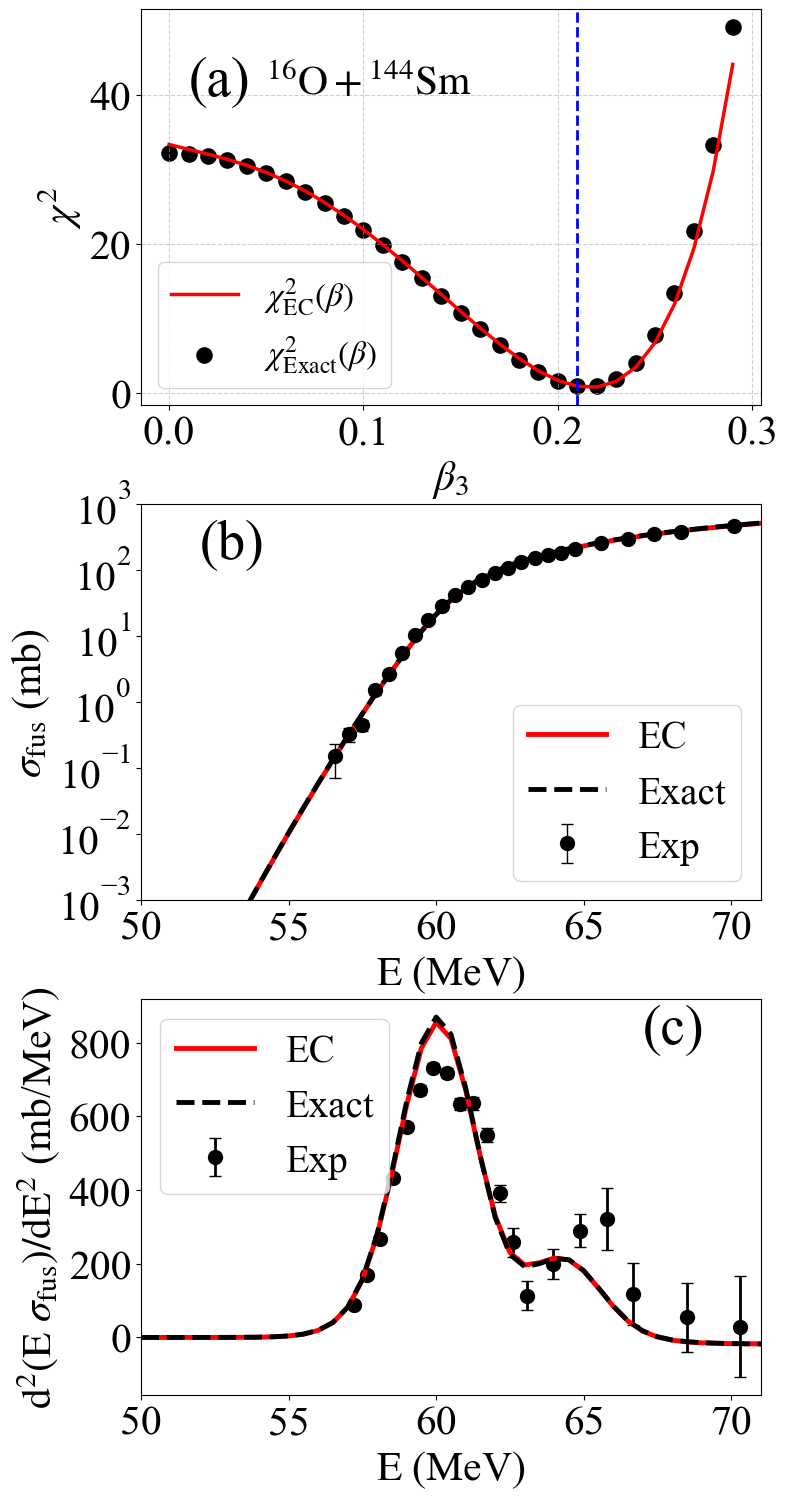}
    \caption{(a) The $\chi^{2}$ function for the $^{16}$O+$^{144}$Sm fusion reaction. It is plotted as a 
    function of the octupole deformation $\beta_3$ of the $^{144}\mathrm{Sm}$ nucleus. 
    The black dots are obtained with the exact fusion cross sections, while 
    the red solid line with the fusion cross sections with the emulator. 
    The blue dashed line indicates the minimum point of the $\chi^{2}$ function. 
 (b) The fusion cross sections $\sigma_{\mathrm{fus}}$ for the $\mathrm{^{16}O}+\mathrm{^{144}Sm}$ reaction as a function of the center-of-mass energy $E$. 
 The black dashed line shows the exact result with the optimum deformation parameter, while the red solid 
 line is obtained with the emulator. The experimental data are taken from 
Ref. \cite{PhysRevC.52.3151}. (c) The corresponding fusion barrier distributions. } \label{Pic_2}
\end{figure}

Following the workflow shown in Sec. \ref{workflow}, 
we apply the emulator to this reaction to extract the deformation parameters of the target nucleus from 
the measured fusion cross sections. 
To this end, we take into account the coupling only to the first 3$^-$ state in $^{144}$Sm. 
We take the parameters of the Woods-Saxon potential, Eq. (\ref{eq:Potenital_nuclear}), to be
$V_{0} = 105.1 \text{ MeV}$, $r_{0} = 1.10 \text{ fm}$, and $a = 0.75 \text{ fm}$.
We discretize the coordinate $r$ from $r_\mathrm{min}$ to $r_\mathrm{max}=30$ fm 
with the mesh spacing of $\Delta r = 0.05 \text{ fm}$. 
For the fusion cross sections, the angular momentum $J$ is taken up to $J_{\rm max}$, 
whose contribution to the cross section is less than $10^{-4}$ times the total fusion cross section.
To determine the optimum deformation parameter $\beta_3$, 
we vary the value of $\beta_3$ in the range of $0.0$ to $0.3$ with a step size of $0.01$, 
and compute the 
$\chi^{2}$ function, 
\begin{equation}
    \chi^2(\beta_3) = \frac{1}{N_{\mathrm{exp}}}\sum_{i=1}^{N_{\mathrm{exp}}} \left(\frac{\sigma_{\rm fus}^{\rm(exp)}(E_i) 
    - \sigma_{\rm fus}(E_i;\beta_3)}{\delta\sigma_{\rm fus}^{\rm(exp)}(E_i)}\right)^2,
    \label{equation_chi}
\end{equation}
where $N_{\mathrm{exp}}$ is the number of the data points, $\sigma_{\rm fus}^{\rm(exp)}(E_i)$ and 
$\delta\sigma_{\rm fus}^{\rm(exp)}(E_i)$ are the experimental fusion cross section and its uncertainty at 
the energy $E_i$, respectively, and $\sigma_{\rm fus}(E_i;\beta_3)$ is the calculated fusion cross section 
at $E_i$ for a given $\beta_3$. 
For simplicity, we take the same value of $\beta_3$ both for the nuclear and the Coulomb couplings. 

Fig. $\ref{Pic_2}$ (a) shows the 
$\chi^{2}$ as a function of the deformation parameter $\beta_3$. 
The black dots show the exact results obtained with the discrete basis formalism, 
while the red solid line represents the results with the emulator. For the latter, we 
take $N_{\mathrm{EC}} = 5$ with $\beta_{3} = 0.05, 0.10, 0.15, 0.20, 0.25$, and apply the eigenvector 
continuation at each $E$ and $J$ separately. 
Both the approaches clearly indicate that the $\chi^2$ minimum appears at $\beta_3 = 0.21$ (see the blue dashed line), 
that is in excellent agreement with the value estimated from the measured $B(E3)$ value. 

Figs. \ref{Pic_2} (b) and (c) show the fusion cross sections and the barrier distributions 
calculated with the optimum value of $\beta_3$, respectively. To extract the barrier distribution from 
the fusion cross sections, we use the point difference formula 
with $\Delta E_{\rm c.m.}=1.8$ MeV \cite{PhysRevC.52.3151}. 
One can see that the exact results (the black dashed line) and the results of the emulation (the red solid line) are almost indistinguishable. This outcome confirms that the emulator can efficiently emulate 
the fusion cross sections across the entire energy range $E$. 
Furthermore, one can see that the calculations well reproduce the experimental fusion cross sections, 
even though some deviation can be seen in the barrier distribution. The shape of the barrier distribution 
can be improved by taking into account the quadrupole excitations in $^{144}$Sm in addition to the 
octupole vibration, as well as the anharmonic effects of these vibrations \cite{PhysRevLett.79.2943}. 

We point out that rigorously speaking 
the optimized $\beta$ values extracted within the present truncated coupled-channels framework 
should be treated as effective values, given that 
the Hamiltonian, Eq. \ref{eq:totH}, for the coupled-channels calculations is 
inherently incomplete. 
Since the main couplings have been taken into account, we do not expect that 
errors in the optimized deformation parameters are large.  
Nevertheless the optimization process would to some extent compensate for model defects, 
such as the omission of explicit transfer channels or higher order deformation effects.  
A rigorous mathematical approach to mitigate this issue would involve 
introducing a model discrepancy term during the calibration process to account for missing physics, 
see e.g., Ref. \cite{kennedy_bayesian_2001}. 
While explicitly modeling this discrepancy term is beyond the scope of the current work, 
it would highlight a critical direction for 
a future uncertainty quantification in heavy-ion fusion analyses.

\subsection{Numerical accuracy of the emulator}

\begin{figure}[htbp]
    \centering
        \includegraphics[width=7.5cm]{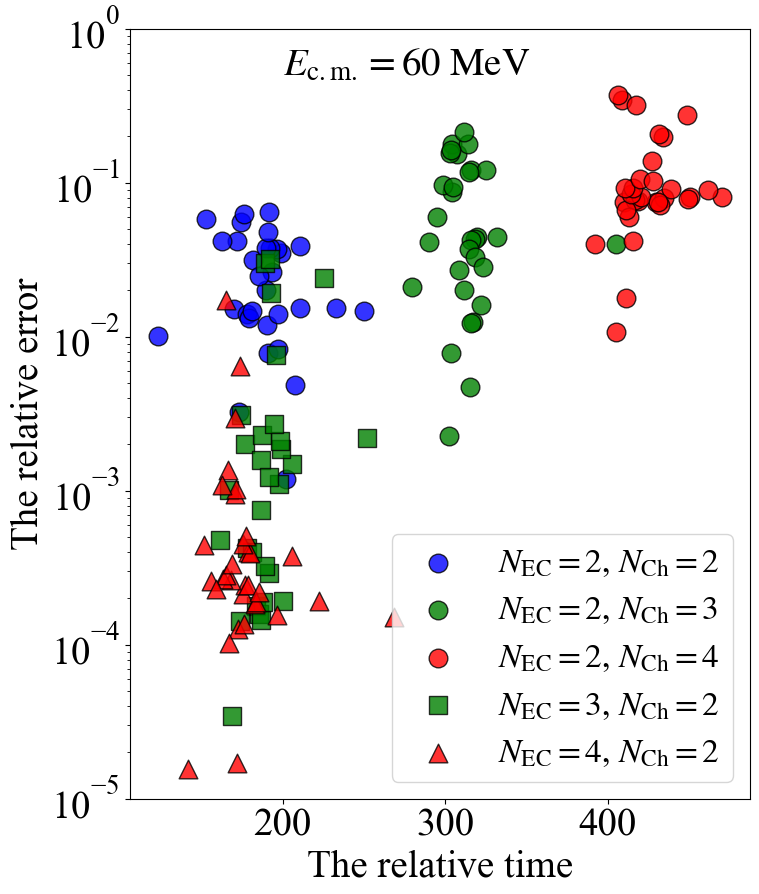}
    \caption{The relative error for the penertrability $P_J$ at the center-of-mass energy $E$=60 MeV between the exact calculations 
    and the EC method as a function of the relative computational consumption. 
Here, the relative error is defined by $|P^{\mathrm{EC}}_{J}(E)-P^{\mathrm{Exact}}_{J}(E)| / P^{\mathrm{Exact}}_{J}(E)$, where $P^{\mathrm{Exact}}_{J}(E)$ and $P^{\mathrm{EC}}_{J}(E)$ are the exact penetration probability and that obtained with the eigenvector continuation, respectively. On the other hand, the relative time is defined by $T^{\mathrm{Exact}}_{J}(E) / T^{\mathrm{EC}}_{J}(E) $, where $T^{\mathrm{Exact}}_{J}(E)$ and $T^{\mathrm{EC}}_{J}(E)$ are the exact comsuming time and that obtained with the eigenvector continuation, respectively. 
Each point is for several combinations of the number of basis state for the EC, $N_{\rm EC}$, and the number of 
channels in the coupled-channels calculations, $N_{\rm Ch}$, and for various values of the angular momentum, $J$. 
} \label{Pic_3}
\end{figure}

The computation efficiency is a critical metric for the emulator. 
To discuss this, 
we plot in Fig. $\ref{Pic_3}$ 
the relative errors of the  penetrability, $P_J$, at a fixed energy of $E=60 \text{ MeV}$ 
as a function of 
the ratio of the computational time between the DBM (Eq. (\ref{eq:Pexact})) and the emulator (Eq. (\ref{ec:P_ec})) for the $\mathrm{^{16}O}+\mathrm{^{144}Sm} $ reaction with $\beta_{3} = 0.21$.
Each point in the figure represents the calculated data for several values of $J$: 
the blue, the green, and the red circles represent the results for different values of 
$N_{\mathrm{ch}}$ but with the same $N_{\mathrm{EC}}$, that is, $N_{\mathrm{EC}}$=2. 
Here, $N_{\mathrm{Ch}}$ corresponds to the number of phonon states to be included in the coupled-channels 
calculations. 
That is, $N_{\mathrm{Ch}} = 2, 3,$ and 4 correspond to couplings up to $1, 2,$ and $3$ octuple phonons, 
respectively. To this end, we assume the harmonic oscillator phonons, as employed in 
{\tt CCFULL} \cite{Hagino:1999xb}. 
On the other hand, 
the blue circles, the green squares, and the red triangles represent the results for 
different values of $N_{\mathrm{EC}}$ but with the same channel number $N_{\mathrm{ch}}$=2. 
For the training sets for EC, we take $\beta_3$ = 0.19 and 0.23 for 
$N_{\mathrm{EC}} = 2$, $\beta_3$ = 0.19, 0.23, and 0.25 for $N_{\mathrm{EC}} = 3$, 
and $\beta_3$ = 0.17, 0.19, 0.23, and 0.25 for $N_{\mathrm{EC}} = 4$. 

A clear trend of rapid convergence in the numerical error can be observed as the number of basis states $N_{\mathrm{EC}}$ increases. Each additional basis state $N_{\mathrm{EC}}$ roughly reduces the deviation 
from the exact result by about an order of magnitude. 
Furthermore, the computational advantage of the EC emulator becomes increasingly significant as 
the complexity of the coupled-channel model, that is, the number of channels $N_{\mathrm{Ch}}$, grows. 
Notice that the dimension of the Hamiltonian matrix is 918, 1377, and 1827 for 
$N_{\mathrm{Ch}}$=2,3, and 4, respectively, thus the computational burden quickly increases as $N_{\mathrm{Ch}}$ 
increases.
For these, the emulator is faster than the exact calculation by a factor of about 200, 300, and 400, respectively. 
In these large-scale calculations, the EC emulator exhibits a pronounced advantage in computational speed compared with the full DBM, confirming its usefulness to solve complex problems, even though its accuracy may slightly 
decrease when maintaining a low value of $N_{\mathrm{EC}}$. 
This dramatic reduction in the computational time is the essential merit for 
building a truly efficient scattering emulator. 

Notice that the relative time shown in Fig. $\ref{Pic_3}$ represents the online speedup achieved during the parameter evaluation process. Constructing the emulator includes an offline cost $T^{\mathrm{offline}}$ \cite{RevModPhys.96.031002}, that is primarily 
the computation of $N_{\mathrm{EC}}$ high-fidelity snapshots. 
Consequently, there exists a computational break-even point. The emulator becomes advantageous when the total number of evaluations $N_{\mathrm{eval}}$ satisfies $N_{\mathrm{eval}}\, T^{\mathrm{exact}} >  T^{\mathrm{offline}} + N_{\mathrm{eval}} \,T^{\mathrm{EC}}$, where $T^{\mathrm{exact}}$ and $T^{\mathrm{EC}}$ are the single-evaluation times for the exact code and the emulator, respectively. 
To account for the offline cost, we have evaluated the computational break-even point. 
For the cases with $N_{\mathrm{EC}} =$ 2 and 4, the EC 
approach becomes computationally more efficient than running the exact code after approximately $N_{\mathrm{eval}} = 5$ evaluations.

\subsection{Static deformation of $^{154}\mathrm{Sm}$}

\begin{figure}[htbp]
    \centering
        \includegraphics[width=9cm]{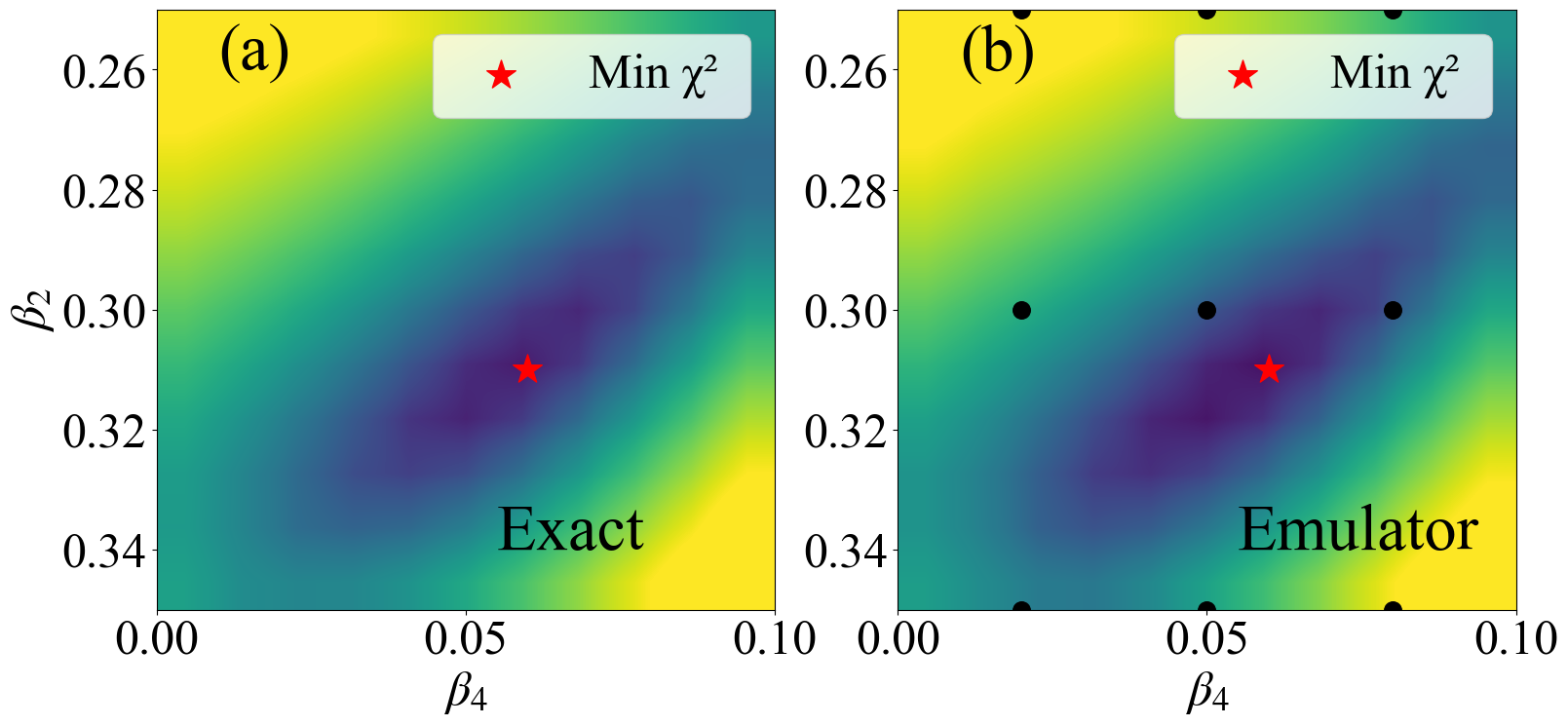}
        \includegraphics[width=9cm]{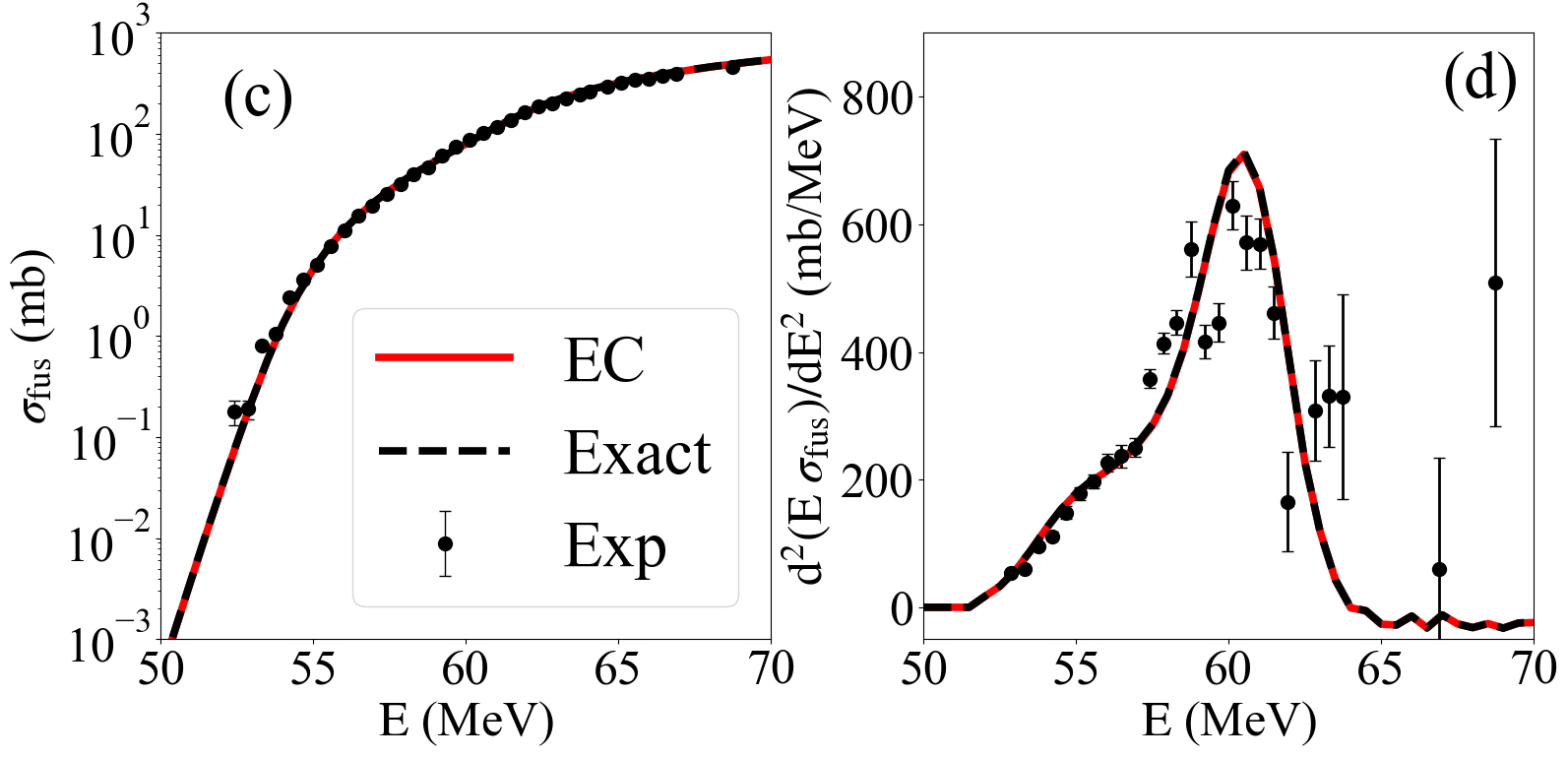}
    \caption{(a) The two-dimensional $\chi^{2}$ function in the ($\beta_2,\beta_4$) plane 
    for the $^{16}$O+$^{154}$Sm fusion reactions, obtained with the exact coupled-channels calculations with the 
    quadrupole deformation parameter $\beta_2$ and the hexadecapole deformation parameter $\beta_4$ of the target 
    nucleus, $^{154}$Sm. The minimum point is denoted by the star. (b) Same as (a), but obtained with the 
    emulator. The black points indicate the training points used for a construction of the emulator. (c) The fusion cross sections, $\sigma_{\mathrm{fus}}$, as a function of the energy $E$ in the center of mass frame for the $\mathrm{^{16}O}+\mathrm{^{154}Sm}$ reaction, calculated with the optimum values of the deformation parameters. The black dashed line shows the exact result, while the red solid line is obtained with the emulator. 
    The experimental data are taken from Ref. \cite{PhysRevC.52.3151}. (d) The corresponding barrier distribution.} 
    \label{Pic_4}
\end{figure}

We next consider the $^{16}\mathrm{O} + ^{154}\mathrm{Sm}$ reaction. 
The target nucleus $^{154}\mathrm{Sm}$ has a clear rotational band, and is considered to be strongly deformed 
in the ground state \cite{PhysRevLett.134.022503, PhysRevC.61.014605, PhysRevLett.123.222502}. 
To analyze the experimental data for the $^{16}\mathrm{O} + ^{154}\mathrm{Sm}$ reaction, 
we adopt a rigid rotor model with axially symmetric quadrupole and hexadecapole deformations. 
In the coupled-channels calculations, we take into account the rotational couplings of $^{154}$Sm up to 
the spin $I_{\mathrm{max}} = 12\hbar$ in the ground state rotational band. 
Following Ref. \cite{PhysRevC.61.014605}, we take $V_{0} = 165 \text{ MeV}$, $r_{0} = 0.95 \text{ fm}$, and $a_{0} = 1.05 \text{ fm}$ for the parameters 
of the Woods-Saxon potential. 
To obtain the optimum values of the deformation parameters, 
we systematically explore the two-dimensional parameter space $(\beta_2, \beta_4)$
by varying $\beta_2$ from $0.25$ to $0.35$ with an increment of $0.01$, and $\beta_4$ from $0.00$ to $0.10$ with an increment of $0.01$. 
The total number of points in this space is 11 $\times$ 11 = 121.
The emulator is constructed with the eigenvector continuation, 
for which we use 9 basis vectors correspond to the training points defined by the grid of $\beta_2$ = 0.25, 0.30, and 0.35, 
and $\beta_4$ = 0.02, 0.05, and 0.08.
These points are shown by the black dots in Fig. \ref{Pic_4} (b). 

Figs. $\ref{Pic_4}$ (a) and (b) show the 
$\chi^{2}$ function in the two-dimensional space of $\beta_2$ and $\beta_4$  
obtained with the exact calculations and with the emulator, respectively. 
One can see that the resulting $\chi^2$ plots are fundamentally similar to each other, and thus 
the emulator can evidently serve as an alternative method to reproduce the exact result. 
The optimal deformation parameters, obtained as the minimum point of the $\chi^2$ function, 
are $\beta_{2} = 0.31$ and $\beta_{4} = 0.06$ both for the exact calculation and for the emulator. 
These values are consistent with the deformation parameters determined with 
other different probes. That is, 
$\beta_{2} =0.317$ and $\beta_{4} =0.070$ were extracted from the analysis of $\alpha$-inelastic scattering, 
for which the original values \cite{APONICK1970367} have been rescaled for the radius parameter \cite{PhysRevC.61.014605}. 
Furthermore, $\beta_{2} = 0.2925(25)$ was extracted from the splitting of the isovector 
giant dipole resonance of $^{154}$Sm \cite{PhysRevLett.134.022503}. 

To rigorously quantify the uniqueness of the fit and the associated uncertainties, 
we have also constructed a likelihood function $L(\mathrm{\beta_2, \beta_4}) \propto \exp\left[-\chi^2(\mathrm{\beta_2, \beta_4})/2\right]$ over a discrete grid of deformation parameters. With a uniform prior, the posterior probability distribution is proportional to the likelihood. 
We have therefore computed the posterior mean and the standard deviation as 
estimates of the parameters and their uncertainties, respectively. 
The exact calculations yield $\beta_2 = 0.31 \pm 0.01$ and $\beta_4 = 0.06 \pm 0.01$, with a strong negative Pearson correlation coefficient of $r = -0.82$. We have found that the 
EC emulator reproduces the same results.

Figs. \ref{Pic_4} (c) and (d) show the fusion cross sections and the corresponding barrier distribution, 
respectively, obtained with the optimum values of the deformation parameters. 
One can see that the result of the emulator (the red solid line) is practically indistinguishable from the 
exact result (the black dashed line), 
and moreover successfully reproduce the experimental data.

\subsection{Static deformation of $^{186}\mathrm{W}$}

\begin{figure}[htbp]
    \centering
        \includegraphics[width=9cm]{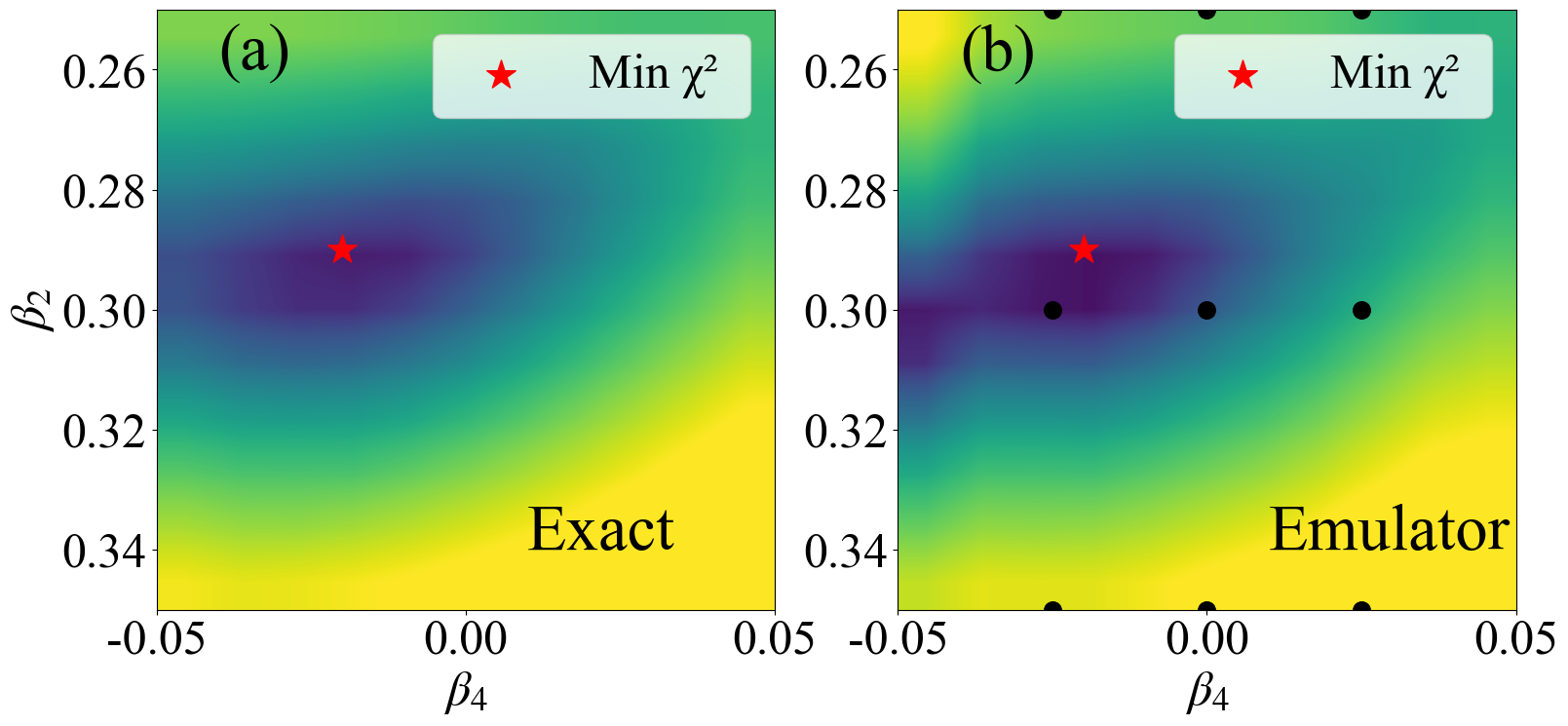}
        \includegraphics[width=9cm]{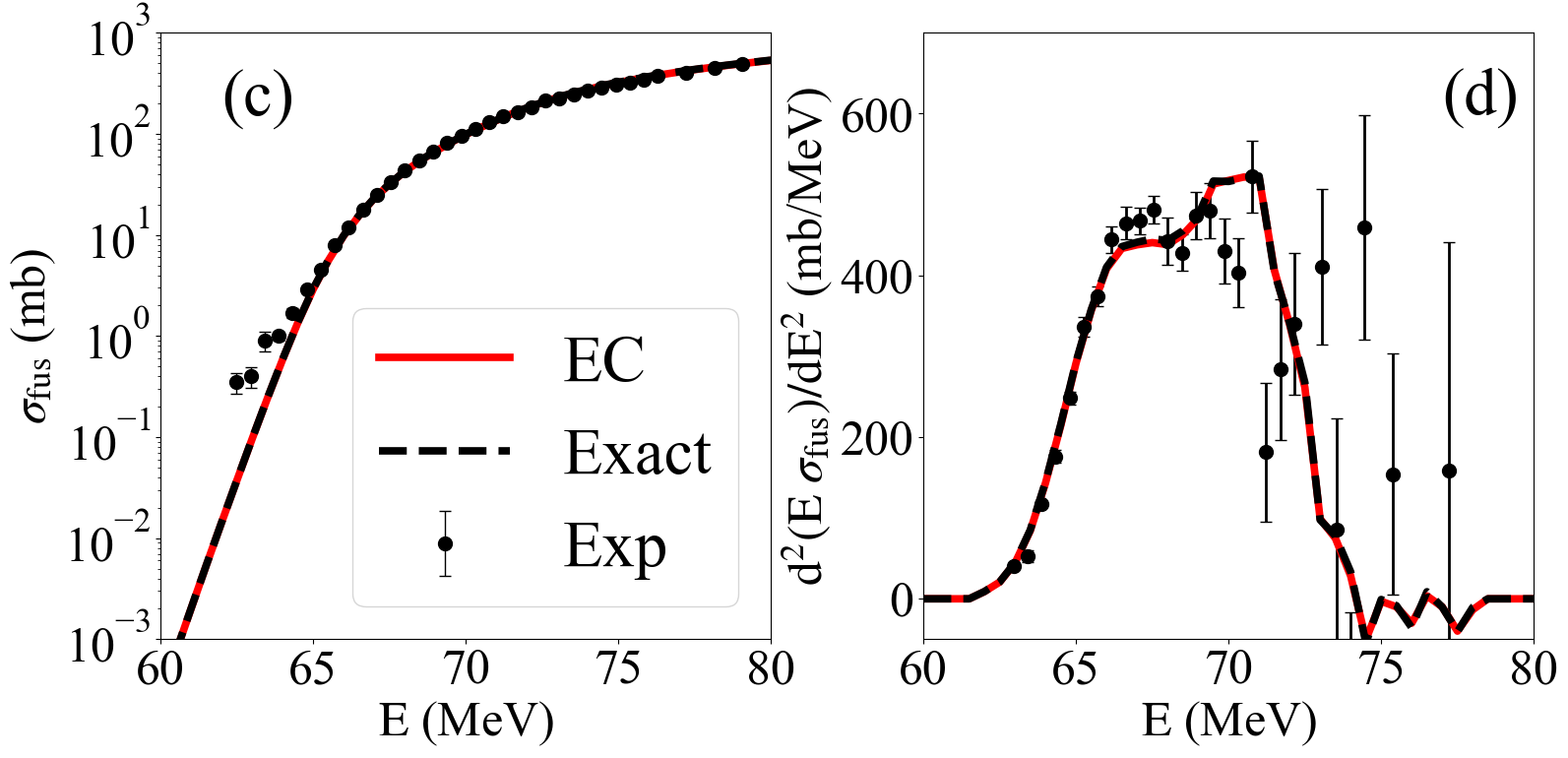}
    \caption{Sames as Fig. \ref{Pic_4}, but for the $^{16}$O+$^{186}\mathrm{W}$ reaction.} \label{Pic_5}
\end{figure}

Following the same procedure, we also analyze the $^{16}$O + $^{186}$W system to determine the deformation parameters of the $^{186}\mathrm{W}$ nucleus. 
To this end, we take $V_{0} = 188 \text{ MeV}$, $r_{0} = 0.95 \text{ fm}$, and $a_{0} = 1.05 \text{ fm}$ for the Woods-Saxon potential, and vary $\beta_2$ and $\beta_4$
in the range of (0.25,0.35) and ($-$0.05,0.05), respectively, with an increment of 0.01. 
We construct an emulator using 9 basis vectors, defined with $\beta_2=$ 0.25, 0.30, and 0.35, 
and $\beta_4=-0.025$, 0.00, and 0.025, as shown by the black dots in Fig. \ref{Pic_5} (b).
The resultant $\chi^2$ function are shown in 
Figs. $\ref{Pic_5}$ (a) and (b) for the exact coupled-channels calculations and for the calculations with the 
emulator, respectively. 
As in the $^{16}$O+$^{154}$Sm shown in the previous subsection, 
the extracted parameters using the emulator are $\beta_2 = 0.29 \pm 0.006$ and $\beta_4 = -0.02 \pm 0.014$, which are consistent with the exact results of $\beta_2 = 0.29 \pm 0.005$ and $\beta_4 = -0.02 \pm 0.011$. 
These value can be compared with the values obtained with earlier neutron scattering, 
$\beta_2=0.203\pm0.006$ and $\beta_4=-0.057\pm0.006$ \cite{PhysRevC.26.1899}. 
Notice that both fusion and neutron scattering yield a negative value of $\beta_4$ of $^{186}$W 
\cite{PhysRevC.52.3151}. 
The fusion cross sections and the corresponding barrier distributions obtained with the optimum 
values of the deformation parameters are shown in Figs. \ref{Pic_5} (c) and (d), respectively. 
Once again, 
one can see that the emulator (the red solid line) successfully reproduce the exact 
calculations (the black dashed line) as well as the experimental data. 

We notice that 
the Pearson correlation coefficients significantly differ between the emulator and the exact 
calculation, even though both of them lead to similar values of the optimum deformation parameters. 
That is, the Pearson correlation coefficient is 
$r = -0.51$ for the emulator, while it is $r = -0.29$ for the exact calculation. 
It is likely that this discrepancy arises from inaccuracies in the predictions with the emulator 
near the boundaries of the parameter space. This would indicate 
that the current basis set used for the emulator may not be optimally chosen. 
The worse performance at the left edges of the grid somewhat 
distorts the joint posterior distribution and overestimates the anti-correlation between $\beta_2$ and $\beta_4$, even though the posterior means and the standard deviations are well reproduced. 
This would be cured to a large extent if one employed 
an algorithmic optimization for snapshot selection.

\section{SUMMARY}\label{sec4}

We have introduced a novel and efficient method for low-energy nuclear reactions based on an emulator integrated with the coupled-channels formalism. To this end, we have employed the eigenvector continuation technique together with the discrete basis formalism. 
We have applied this to 
the $^{16}$O+$^{144,154}$Sm, $^{186}$W fusion reactions at energies around the Coulomb barrier, and 
demonstrated that the emulator accurately reproduces key observables of the reactions while achieving a substantial speed-up over 
the conventional methods. 
We have also demonstrated that the emulator can successfully lead to optimum values of the deformation parameters which are 
consistent to the values with the exact calculations. This was the case both for the surface vibration of ${^{144}\mathrm{Sm}}$ and 
the static deformation of ${^{154}\mathrm{Sm}}$ and ${^{186}\mathrm{W}}$. 
This validates the capacity of the emulator to accurately extract intrinsic nuclear properties. 
The observed rapid convergence of numerical errors, along with the substantial speed-up over the exact 
calculations, confirms EC as an efficient and reliable surrogate model for coupled-channels calculations. 
Given these superior characteristics of EC, the EC approach is ideally suited for computationally 
intensive applications involving numerous repeated calculations, particularly multi-parameter searches. 
This methodological advancement offers a systematic way to explore the shapes of atomic nuclei.  
Ultimately, the emulator introduced in this paper will enhance our ability to precisely determine the fundamental properties of 
atomic nuclei, paving the way for more comprehensive and accurate reaction studies.

In this paper, for the deformed nuclei, we have considered axial quadrupole and hexadecapole deformations 
within the coupled-channels formalism. 
We mention that 
the theoretical framework can be readily extended to encompass more complex deformations, e.g., 
the triaxial ($\gamma$) and the octupole ($\beta_3$) deformations. 
Given the current intense interest in the high-energy 
nuclear physics community regarding a determination of these deformations 
of e.g., $^{238}\mathrm{U}$ \cite{star2024imaging, Collaboration_2025}, 
we plan to explore optimal deformation parameters of this nucleus applying the theoretical basis of the present study 
to e.g. $^{16}$O+$^{238}$U fusion reactions at subbarrier energies \cite{PhysRevLett.74.1295,PhysRevLett.93.162701}. 
Furthermore, to manage the increased dimensionality of the parameter space and further enhance the efficiency of emulators, it would be useful to incorporate an algorithmic optimization for training snapshot selection\cite{ChenChen2017} and to utilize the Principal Component Analysis \cite{quarteroni2016reduced} in order 
to construct a more compact reduced basis.
We will report on these in a separate publication.

\appendix  

\section{Discrete basis formalism with Numerov method}\label{appendix:A}

\begin{figure}[htbp]
    \centering
        \includegraphics[width=8cm]{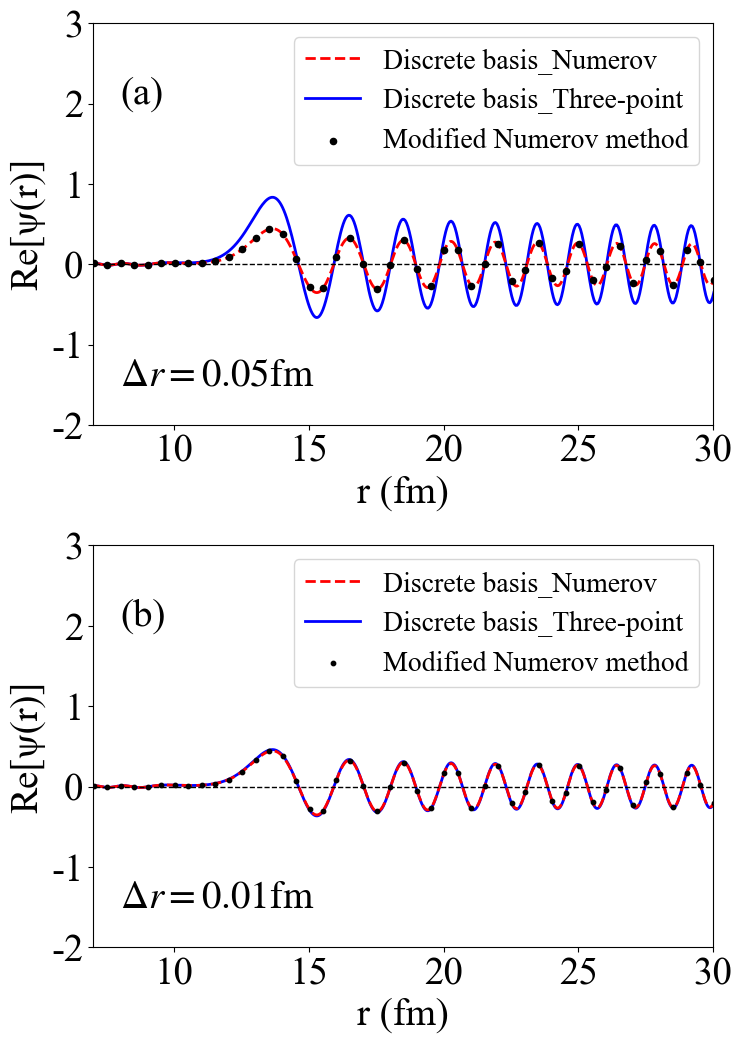}
    \caption{(a) A comparison of several numerical methods for the wave function for the first channel 
    in the $s$-wave $^{16}\mathrm{O} + \mathrm{^{144}Sm}$ scattering at $E = 55 \text{ MeV}$. 
    The mesh size is taken to be $\Delta r = 0.05$ $\mathrm{fm}$.
The black dots represent the results of the modified Numerov algorithm. 
The blue solid line shows the results of the discrete basis formalism with the three-point formula, while 
the red dashed line is for those with the Numerov method. (b) Same as (a), but calculated with a 
smaller mesh spacing of $\Delta r = 0.01$ $\mathrm{fm}$.}  \label{Pic_6}
\end{figure}

The coupled-channels equations, Eqs. (\ref{eq:cc}) and (\ref{eq:cc-disc}), can be solved 
with several numerical methods. 
The modified Numerov methods is adopted in the computer code {\tt CCFULL} \cite{Hagino:1999xb}, while 
the discrete basis formalism in Refs. \cite{zpsz-7jld,PhysRevC.110.054610} used the three-point formula for the 
kinetic energy operator, with which 
the Hamiltonian can be transformed into the matrix form as in Eq. (\ref{eq:H}). 

In the discrete basis formalism, 
the computation cost and the convergence of the scattering wavefunction 
depends on the value of mesh spacing, $\Delta r$. 
If $\Delta r$ is chosen to be too large, the performance of the emulator 
constructed with such wave functions is poor. 
Conversely, if $\Delta r$ is too small, the dimension of the Hamiltonian matrix, Eq. (\ref{eq:H}), 
becomes exceptionally large, making the calculation extremely time-consuming.

In order to enhance the convergence without increasing the computational load, in this paper, we utilize the modified Numerov 
method for the kinetic energy operator, instead of the three-point formula. 
In this method, the coupled-channels equations are transformed to \cite{koonin,Hagino:1999xb}
\begin{equation}
\begin{aligned}
  &  \chi_{n,i-1} + \chi_{n,i+1} - \\
  &  \sum_{n'}\left\{ \left(\frac{\Delta r^{2}}{\sqrt{12}}\vec{A} + \sqrt{3}\cdot \vec{1} \right)^{2} -\vec{1}\right\}_{ni,n'i} \chi_{n'i}  = 0,\label{eq:numerov}
\end{aligned}
\end{equation}
where \vec{1} is the unit matrix and $\chi_{ni}$ is defined by 
\begin{equation}
    \chi_{ni} = \sum_{n'}\left(\vec{1} - \frac{\Delta r^{2}}{12} \vec{A}\right)_{ni,n'i} \phi_{n'i},
\end{equation}
with the matrix $\vec{A}$ defined as 
\begin{equation}
\begin{aligned}
 A_{ni,n'i'} &= \frac{2\mu}{\hbar^2} \left(  \frac{J(J+1)\hbar^2}{2\mu r_{i}^2} + V^{0}(r_{i})+ \epsilon_n - E \right) \delta_{n,n'} \delta_{i,i'} \\
  &+ \frac{2\mu}{\hbar^2} V_{nn'}(r_{i})  \delta_{i,i'}. 
\end{aligned}
\end{equation}
Here, $\phi_{ni}$ is the wave function defined in Ref. (\ref{eq:cc-disc}). 
With this, the Hamiltonian matrix reads, 
\begin{equation}
\begin{aligned}
&H_{ni,n'i'}=\left(\delta_{n,n'}\delta_{i,i'} - \frac{\Delta r^{2}}{12} A_{ni,n'i'}\right) \\ & 
\times\left(\delta_{i,i'+1} - \left\{ \left(\frac{\Delta r^{2}}{\sqrt{12}}A_{ni,n'i'} + \sqrt{3}\cdot \delta_{n,n'}\right)^{2} -\delta_{n,n'} \right\}\,\delta_{i,i'} \right. \\ & 
+ \delta_{i,i'-1}). 
\end{aligned}
\end{equation}
The modified Hamiltonian that corresponds to Eq. (\ref{eq:H-disc}) reads, 
\begin{equation}
\begin{aligned}
\tilde{H}_{ni,n'i'} &= H_{ni,n'i'} \\
& +\left(\delta_{n,n'} - \frac{\Delta r^{2}}{12} A_{ni,n'i'}\right)e^{ik_n(r_{\rm min})\Delta r}\delta_{i,i'}\delta_{i,1}. 
\end{aligned}
\end{equation}
Notice that the energy $E$ is already included in the matrix \vec{A}. 
Once the Hamiltonian matrix is so constructed, all the subsequent steps remain the 
same as those in Sec. \ref{subsec:DB}. 

Fig. \ref{Pic_6} (a) compares 
the direct integral of the coupled-channels equations wih the method with modified Numerov method (the dots), 
the discrete basis formalism with the Numerov method (the dashed line), and that with the three-point 
formula (the solid line) for the 
the real part of wave function for the first channel obtained with the mesh size of 
$\Delta r = 0.05$ $\mathrm{fm}$. 
To this end, we consider 
the $s$-wave $^{16}\mathrm{O} + \mathrm{^{144}Sm}$ scattering at $E = 55 \text{ MeV}$. 
One can see that the result with the three-point formula deviate from the results with the other methods. 
Only with a smaller mesh size, such as, $\Delta r = 0.01$ $\mathrm{fm}$, the result of the 
three-point formula converges to the results of the other methods, as shown in Fig. \ref{Pic_6} (b). Obviously, 
the discrete basis formalism with the Numerov method 
provides a more efficient method without increasing the matrix dimension. 
We find that the results of the discrete basis formalism with the modified Numerov method do not significantly change even if $\Delta r$ is increased 
to $\Delta r$ = 0.1 fm, even though the deviation from the converged result becomes significant for 
$\Delta r$ = 0.2 fm.

For a mesh spacing of $\Delta r = 0.05$ $\mathrm{fm}$, the size of the Hamiltonian matrix is $918 \times 918$, while the size increases to $4594 \times 4594$ for $\Delta r = 0.01$ $\mathrm{fm}$. 
Even though the convergence is achieved with the three-point formula with $\Delta r = 0.01$ $\mathrm{fm}$ shown as Fig. \ref{Pic_6}.(b),  
the significant increase in the matrix size leads to a vast computational burden, which is further amplified 
when the number of channels increases. Consequently, the choice of $\Delta r = 0.05 \text{ fm}$ in the present work strikes a balance between computational efficiency and numerical accuracy.

\begin{acknowledgments}
The authors thank X. Zhang, Xiaoyu Wang, and M. Kimura for useful discussions. Z.L. also thanks 
the nuclear theory group of the Kyoto University for its hospitality. 
This work was in part supported by JSPS KAKENHI Grant Number JP23K03414
JP22K14030, JP25H01511; JST PRESTO (Grant No.~JPMJPR25F8); JST ERATO (Grant No.~JPMJER2304);
RIKEN TRIP initiative (Nuclear Transmutation);
International Program for Candidates, Sun Yat-sen University; The National Natural Science Foundation of China under Grant No. 12075327.
\end{acknowledgments}

\medskip

\begin{center}
{\bf DATA AVAILABILITY}     
\end{center}

The data that support the findings of this article are available from the authors upon reasonable request.

\bibliography{apssamp}

\end{document}